\documentstyle[aps,preprint,12pt]{revtex}

\begin{document}
\large

\title{\Large \bf On U(1)-charged domain walls}

\author{V. A. Lensky \thanks{E-mail: lensky@vxitep.itep.ru}}

\address{ {\it Moscow State Engineering Physics Institute,}\\
{\it Russia, 115409, Moscow, Kashirskoe shosse, 31}\\
}

\author{V. A. Gani \thanks{E-mail: gani@heron.itep.ru},
A. E. Kudryavtsev \thanks{E-mail: kudryavt@heron.itep.ru}}

\address{ {\it Institute of Theoretical and Experimental Physics,}\\
{\it Russia, 117259, Moscow, Bolshaya Cheremushkinskaya st., 25}\\
}

\maketitle

\begin{abstract}
A classical field system of two interacting fields -- a real Higgs
field and a complex scalar field -- is considered. It is shown that
in such field system a non-trivial solution exists, which is U(1)
charged topological kink. Some questions of stability of the obtained
solution are discussed. An improved variational procedure for searching
of topological U(1) charged solutions is given.
\end{abstract}

\section{Introduction}

     It is well-known that solutions of the type of domain walls are
of great importance in a wide class of modern supersymmetric
field theories and in string description (D-branes) of gauge theories
\cite{Kovner}. A problem of interaction of particles with domain walls 
has been discussed widely starting from the paper of Voloshin \cite{Vol1}.
In Ref. \cite{Farrar} the interaction between abelian gauge particle 
and domain walls has been studied in detail. Usually, the interaction 
between a particle and a wall is being considered as a scattering in a 
given potential, which is created by wall, i.e. in the approximation 
of a weak coupling, and the particle influence on the wall is neglected.
Such approach may be sometimes incorrect.
For example, it is known that the interaction between skyrmions
or monopoles and domain walls is non-trivial \cite{Dvali,Zakrzewski}.
Besides that, walls are able to form bound states
with other fields and to carry quantum numbers of
these fields. Recently, an integrable dyonic-type lattice model
with U(1) charged topological soliton has been considered \cite{Gomes}.

      In the present paper we will discuss a simple continuous model for 
the system of two interacting scalar fields. In this model, existence of 
non-topological U(1) charged solutions (Q-balls) was approved in Ref. 
\cite{Lee}. As it will be shown in this our work, in (1+1) dimensions in the 
Friedberg-Lee-Sirlin model kink-like topological solutions exist, which
in addition carry U(1) charge. Notice that monopole-like solutions,
carrying also electric charge, are well-known. Another example of U(1)
charged topological solution is the so-called Q-lump \cite{Leese}.

\section{Topological and non-topological Q-balls}

   Let us consider the system of two interacting scalar fields     
in $(1+1)$ dimensions with the Lagrangian density
$$
{\cal L}=\partial_\mu\xi\partial^\mu\xi^*+\frac{1}{2}\partial_\mu\phi
\partial^\mu\phi-h^2\phi^2\xi\xi^*-\displaystyle\frac{m^2}{2}(\phi^2-v^2)^2.
\eqno(1)
$$
Here $\phi$ is a real scalar field, $\xi$ is a complex scalar field;
$h, m, v$ are real and positive constants; $\mu=0, 1$.
Such field Lagrangian in (3+1) dimensions was examined for the first time in work
 \cite{Lee}. In particular, in \cite{Lee} classical solutions of Euler's
equations for the Lagrangian (1), having the spherical symmetry and 
U(1) charge, were obtained. Later, non-topological soliton solutions 
similar to these
were called "Q-balls" \cite{Coleman}. One-dimensional Q-balls for 
a field system close to (1) are considered in detail in book \cite{Rubakov}.

     Lagrangian density (1) is invariant under global U(1) transformations
$$
\xi\to\xi e^{i\alpha},
\eqno(2)
$$	
and also under discrete transformation
$$
\phi\to-\phi.
\eqno(3)
$$
Symmetry (2) generates the conserving current
$$
j^\mu=\frac{1}{i}(\xi^*\partial^\mu\xi-\xi\partial^\mu\xi^*)
\eqno
$$
and charge
$$
Q=\int j^0 dx.
$$
Symmetry (3) is responsible for the conserving topological current
$$
i^\mu=\epsilon^{\mu\nu}\partial_\nu\phi
\eqno
$$
and topological charge
$$
P=\int i^0 dx
$$
(here $\epsilon^{\mu\nu}$ is the Levi-Civita tensor).

     Lagrangian (1) implies the equations of motion
$$
\partial_\mu\partial^\mu\xi+h^2\phi^2\xi=0,
\eqno(4)
$$
$$
\partial_\mu\partial^\mu\phi+2h^2\xi\xi^*\phi+2m^2(\phi^2-v^2)\phi=0.
\eqno(5)
$$
The vacua of system (1) are determined as
$$
\begin{tabular}{l}
$\phi_{vac}=\pm v$,\\
$\xi_{vac}=0$.
\end{tabular}
\eqno(6)
$$
If we are interested in such field configuration that the field $\xi$
is equal to its vacuum value, then equation (4) is satisfied,
and (5) takes the form
$$
\partial_\mu\partial^\mu\phi+2m^2(\phi^2-v^2)\phi=0,
\eqno(7)
$$
which is identical to the equation of motion for the $\lambda\phi^4$ 
theory with two degenerate vacua.
In (1+1) dimensions Eq. (7) has a static topological solution --
"kink":
$$
\phi_k(x)=v\tanh(mvx).
\eqno(8)
$$
Because of $\xi=0$ in the configuration being considered,
the U(1) charge is also equal to zero. Energy of this
configuration (the mass of kink) is
$$
E_{k}=\frac{4}{3}mv^3.
\eqno(9)
$$

    Next, we will show that in field system defined by the Lagrangian
(1), at large charge $Q$ there exist localised configurations that are stable
with respect to decay into plane waves. Such configurations could be
either non-topological (Q-balls) or topological (topological Q-balls).
Hereafter, we will need the expression for the energy of plane waves,
that are the excitations of fields over the vacua (6). By that, for
avoiding of divergent values in energy and charge calculation, we
suppose that these excitations exist in large but finite segment of $X$
axis: $-L/2\le x\le L/2$. Supposing amplitudes of deviations of fields
from (6) to be small, we get in linear approximation from (4) and (5):
$$
\begin{tabular}{l}
$\partial_\mu\partial^\mu\delta\xi+h^2v^2\delta\xi=0$,\\
\\
$\partial_\mu\partial^\mu\delta\phi+4m^2v^2\delta\phi=0$,
\end{tabular}
\eqno(10)
$$
where $\delta\phi=\phi-\phi_{vac}, \delta\xi=\xi-\xi_{vac}$.
At fixed charge of configuration the solution of (10), that minimizes 
the deviation of energy from its vacuum value, is
$$
\begin{tabular}{l}
$\delta\phi=0$,\\
$\delta\xi=\sqrt{\displaystyle\frac{Q}{2hvL}}\:e^{ihvt}$.
\end{tabular}
\eqno(11)
$$
The energy of solution (11) at given charge $Q$ is
$$
E^0=hvQ.
\eqno(12)
$$

     The essential difference between the cases of one and three 
space dimensions is the existence in (1+1) dimensions in system (1)
of topological solutions. Considering the possibility of existence
of stable topological charged solutions we have to compare the energy
of such solution, having the charge $Q$, with the sum of the energy (9)
of the kink (8) (which has the lowest energy among all topological 
configurations) and the energy of charged non-topological configuration (11).
The value of this sum is 
$$
\tilde{E^0}=hvQ+\frac{4}{3}mv^3.
\eqno(13)
$$ 
Hereafter, we will use this value as starting point for searching 
of possibility of existence of a stable charged topological configuration.

     Now, following \cite{Lee,Rubakov}, let us look at the energy functional
$$
H[\phi,\xi]=\int dx\left(\left|\frac{\partial\xi}{\partial t}\right|^2+
\left|\frac{\partial\xi}{\partial x}\right|^2+
\frac{1}{2}\left(\frac{\partial\phi}{\partial t}\right)^2+
\frac{1}{2}\left(\frac{\partial\phi}{\partial x}\right)^2
\right.
$$
$$
\left.
+h^2\phi^2\xi\xi^*+\displaystyle\frac{m^2}{2}(\phi^2-v^2)^2
\displaystyle\right)
\eqno(14)
$$
at the class of trial functions
$$
\phi(x)=\left\{
\begin{array}{ll}
0, & |x|<l/2,\\
\pm v\left[1-\exp\left(\displaystyle\frac{x+l/2}{a}\right)\right], & x\le -l/2,\\
v\left[1-\exp\left(-\displaystyle\frac{x-l/2}{a}\right)\right], & x\ge l/2,\\
\end{array}
\right.
\eqno(15)
$$
$$
\xi(x,t)=\left\{
\begin{array}{ll}
0, & |x|>l/2,\\
\\
A\cos(\Omega x)\exp(i\Omega t), & |x|\le l/2,\\
\end{array}
\right.
\eqno(16)
$$
where $l$ and $a$ are arbitrary positive constants which are to vary
(the Ritz parameters). The sign $+$ or $-$ in (15) is chosen
respectively for non-topological and topological trial functions. 
We are interested in the state having the lowest energy.
For this state, $|\xi(x)|$ does not turn to zero at $|x|<l/2$,
so we get
$$
\Omega=\displaystyle\frac{\pi}{l}.
\eqno
$$
In the case $\xi(x,t)=|\xi(x)|\exp(i\Omega t)$, the charge of 
configuration is
$$
Q=2\Omega\int |\xi(x)|^2 dx.
\eqno
$$
Using that, we get the value of $A$ in (16):
$$
A=\displaystyle\sqrt{\frac{Q}{\pi}}.
\eqno
$$
Now, we are able to write the expression for the energy of trial configuration
(15)-(16):
$$
E=\displaystyle\frac{\pi Q}{l}+\frac{v^2}{2a}+\frac{m^2v^4l}{2}+
\frac{11am^2v^4}{12}.
\eqno(17)
$$
The coupling constant $h$ does not appear in expression (17) in accordance to
the fact that it enters in the Hamiltonian only in combination
$h^2\phi^2\xi\xi^*$, but for functions (15)-(16) this product is
equal to zero. Note also that Eq. (17) is of sense both for non-topological
and for topological configurations because of quadratic entering of $\phi$
into the Lagrangian so that sign changing in Eq. (15) does not affect on
the energy value.

     The minimal value of (17) at given charge $Q$
is getting at Ritz parameter values 
$$
\begin{tabular}{l}
$l_0=\displaystyle\frac{\sqrt{2\pi Q}}{mv^2}$,\\
\\
$a_0=\displaystyle\frac{1}{mv}\sqrt{\frac{6}{11}}$
\end{tabular}
\eqno
$$
and is equal to
$$
E_{min}=mv^2\sqrt{2\pi Q}+mv^3\sqrt{\frac{11}{6}}.
\eqno(18)
$$
As a plane wave energy depends on charge linearly (see (12)-(13)), 
then at large charges, the localised configuration (15)-(16),
whose energy depends on $Q$ as $\sqrt{Q}$, is more energetically
favorable. It means that there is a localised solution in the system
(1) at large values of charge $Q>Q^{cr}$. Note that the value of energy 
(13) for topological configuration "kink+plane wave" at given value of $Q$ 
is different from the value of energy (12) for non-topological
plane wave configuration at the same value of $Q$, while the variational
estimation (18) is applicable both to non-topological and topological
configurations. So, we get different values for the critical charge 
$Q^{cr}$.
\footnote{These values are upper bounds obtained by using of trial functions
(15)-(16), so there could be stable charged localised solutions at values of 
charge, smaller than (19)-(20).}\\
1) Non-topological Q-ball:
$$
Q^{cr}=\frac{m^2v^2}{h^2}\left(\pi+\frac{h}{m}
\sqrt{\frac{11}{6}}+\sqrt{\pi^2+2\pi\frac{h}{m}\sqrt{\frac{11}{6}}}\right).
\eqno(19)
$$
2) Topological Q-ball:
$$
\tilde{Q}^{cr}=\frac{m^2v^2}{h^2}\left(\pi+\frac{h}{m}\left(
\sqrt{\frac{11}{6}}-\frac{4}{3}\right)+\sqrt{\pi^2+2\pi\frac{h}{m}
\left(\sqrt{\frac{11}{6}}-\frac{4}{3}\right)}\right).
\eqno(20)
$$
The meaning of critical charges $Q^{cr}$ and $\tilde{Q}^{cr}$
is explained in Fig.~1. So, we get that the values of charge at which
configuration becomes energetically more preferable and consequently stable with respect to
decay in "kink+plane wave", are smaller for the topological configuration
than for the non-topological one.      

\section{Exact solution for topological Q-ball}

    In the previous section, we have shown that in the system with
Lagrangian (1), besides an ordinary Q-ball, exists a topological Q-ball,
having both topological and U(1) charges. However, the solution given in
the previous section is variational. An exact solution of the problem
satisfying given boundary conditions at spatial infinity (being in a
given topological sector) and having given U(1) charge $Q$ may have
smaller energy. As it will be shown in this section, a particular
solution of Eqs. (4)-(5) can be found in analytical form.

     Indeed, let us suppose the solution of Eqs. (4)-(5) to be in the form
$$
\begin{tabular}{l}
$\phi(x,t)=\phi(x)$,\\
\\
$\xi(x,t)=f(x)e^{i\omega t}$,
\end{tabular}
\eqno(21)
$$
where $\phi(x)$ and $f(x)$ are real functions of $x$. Because of Eq. (21),
Eqs. (4) and (5) become
$$
-\omega^2f-f^{''}+h^2\phi^2f=0,
\eqno(22)
$$
$$
-\phi^{''}+2h^2f^2\phi+2m^2(\phi^2-v^2)\phi=0.
\eqno(23)
$$
Then, we assume that the field $\phi$ holds the functional form of the 
kink (8), maybe having other scale:
$$
\phi(x)=v\tanh(\alpha mvx),
\eqno(24)
$$ 
where $\alpha>0$ is an unknown constant.

     Let us turn in Eqs. (22) and (23) to the new independent variable 
$z=v\tanh(\alpha mvx)$, $-v<z<v$. Instead of (22) and (23) we get
$$
-\omega^2f-\alpha^2m^2(v^2-z^2)\frac{d}{dz}\left((v^2-z^2)\frac{df}{dz}\right)
+h^2\phi^2f=0,
\eqno(25)
$$
$$
-\alpha^2m^2(v^2-z^2)\frac{d}{dz}\left((v^2-z^2)\frac{d\phi}{dz}\right)
+2h^2f^2\phi+2m^2(\phi^2-v^2)\phi=0.
\eqno(26)
$$
Taking into account that in accordance with (24), $\phi(z)=z$,
we get from (26):
$$
f^2(z)=\frac{m^2(1-\alpha^2)}{h^2}(v^2-z^2),
\eqno(27)
$$
from where we can see that solution we are looking for exists only if
constant $\alpha$ is smaller than one. Assuming this condition to be 
satisfied, we get after taking square root of (27) and substituting result
expression for $f(z)$ into Eq. (25):
$$
(2\alpha^2m^2-h^2)z^2+(\omega^2-\alpha^2m^2v^2)=0.
\eqno(28)
$$
As (28) has to take place at arbitrary $z$, then we get
$$
\alpha^2=\frac{h^2}{2m^2},
\eqno
$$
$$
\omega^2=\alpha^2m^2v^2=\frac{h^2v^2}{2}.
\eqno
$$
So, we have found an exact non-trivial solution of Eqs (4)-(5). It is:
$$
\phi(x)=v\tanh\left(\displaystyle\frac{hv}{\sqrt{2}}x\right),
\eqno(29)
$$
$$
\xi(x,t)=v\displaystyle\sqrt{\frac{m^2}{h^2}-\frac{1}{2}}\:
\frac{\exp\left(i\displaystyle\frac{hv}{\sqrt{2}}t\right)}{\cosh\left(\displaystyle\frac{hv}{\sqrt{2}}x
\right)}.
\eqno(30)
$$
This solution exists at the following range of the Lagrangian parameters:
$$
\frac{h^2}{m^2}\equiv\rho^2\le 2.
\eqno(31)
$$
Note that the spatial scales of $\phi$ (29) and $\xi$ (30) are
the same, and at $\rho\to\sqrt{2}$
the solution (29)-(30) turns into non-charged topological kink (8). 
Let us write expressions for energy and charge of (29)-(30):
$$
E=\frac{2\sqrt{2}}{3}hv^3\left(4\frac{m^2}{h^2}-1\right),
\eqno(32)
$$   
$$
Q=2v^2\left(2\frac{m^2}{h^2}-1\right).
\eqno(33)
$$

     Now, we will consider changing of the obtained solution
and of the values of its energy and charge with changing of the Lagrangian
parameters. At $h\ll m$ the solution (29)-(30) has the spatial
scale greater than one of the kink (8); at the same time
the magnitude of charged field is large ($\sim v\displaystyle\frac{m}{h}$).
As it follows from Eqs. (32) and (33), the energy and the charge of the exact
solution are also large in this limit of small coupling constant $h$.
The fact that $\xi$ field's magnitude in (30) tends to infinity at $h\to 0$
means that the solution (29)-(30) is non-perturbative with respect to
the parameter $h$.
It can not be obtained by expansion in powers of $h$ in the small $h$ limit
(at $h\to 0$ the interaction term in (1) is of order
of $m^2v^4$, i.e. it is not small in comparison to the others).
On the contrary, the limit of weak coupling in (29)-(30) is the case
when $h\to\sqrt{2}m$. In this limit the magnitude of $\xi$ in (30),
along with the charge, goes to zero, meanwhile $\phi(x)$
approaches $\phi_{k}(x)$ (8). By this, the weak coupling limit 
corresponds not to small $h^2$, but to small $h^2|\xi|^2$.
We note here that the exact topological solution (29)-(30) has an essential
dependence on the constant $h$ in contrast to the variational one, 
which does not have such dependence. 

     Let us compare now the values of energy of the configuration (29)-(30),
of the plane wave configuration, and of the trial functions (15)-(16). To do
this, we should substitute the value (33) of charge $Q$ into (13) and (18).
After doing that we get
$$
\tilde{E^0}=mv^3\left(\frac{4}{3}+2\rho\left(\frac{2}{\rho^2}-1\right)\right),
\eqno(34)
$$
$$
E_{min}=mv^3\left(2\sqrt{\pi\left(\frac{2}{\rho^2}-1\right)}+
\sqrt{\frac{11}{6}}\right).
\eqno(35)
$$
It is not difficult to prove that both the value (34) of the energy of
"kink+plane wave" configuration and the energy (35) of the variational
configuration (we, of course, suppose the charge of all configurations
to be equal to one of the solution (29)-(30)) are greater than the 
value (32) of energy of the exact solution at arbitrary values of 
the Lagrangian parameters satisfying inequality (31).
The fact that the charge $Q$ (33) is less than
the estimation (20) for the critical charge of the topological configuration,
is quite natural because (29)-(30) is the exact solution of equations of motion.
That is why the exact solution (29)-(30) is stable with respect to decay
both into "kink+plane wave" and into configurations, close to the
variational one (15)-(16).

 The dependences of the ratio $E/m$ versus
$\rho$ for the exact solution, configuration "kink+plane wave"
and for the variational configuration (15)-(16) are presented
in Fig.~2 for $0<\rho\le\sqrt{2}$.

\section{Improved variational procedure}

    The existence
of the exact solution (29)-(30) with energy value
smaller than one of the variational configuration
of section II,
implies that the variational procedure we used above should be improved.
Besides that, at given Lagrangian parameters the
charge value (33)
of the exact solution is fixed and as a result,
the exact solution
does not reproduce all the spectrum of topological
solutions with different
values of $Q$.
Let us try
to improve the variational procedure. For this aim,
we choose the trial functions as follows:
$$
\phi(x)=v\tanh{(\beta x)},
\eqno(36)
$$
$$
\xi(x,t)=Av\displaystyle\frac{\exp{(i\beta t)}}{\cosh{(\beta x)}},
\eqno(37)
$$
where $A$ and $\beta$ are the Ritz parameters. This choice of the variational
function at $A=\displaystyle\sqrt{\frac{m^2}{h^2}-\frac{1}{2}}$ and
$\beta=\displaystyle\frac{hv}{\sqrt{2}}$ returns us to the
exact solution (29)-(30).
Meanwhile, one may hope that because of continuous dependence
of the trial functions on parameters $A$ and $\beta$, the substitution of
(36)-(37) into the energy functional (14) will lead to better
minimization
of this functional at least in the range of parameters $A$ and $\beta$,
close to those of the exact solution. Substituting
(36)-(37) into (14),
we get following expressions for energy and charge:
$$
\tilde{E}=\displaystyle\frac{2}{3}v^2\left(1+4A^2\right)\beta+
\frac{2}{3}v^4\left(h^2A^2+m^2\right)\frac{1}{\beta},
\eqno(38)
$$
$$
Q=4A^2v^2.
\eqno(39)
$$
Thus, the value of $A$ is fixed by the choice of charge value.
Putting obtained from Eq. (39) value of $A$ into (38) and
minimizing the energy with $\beta$ at given $Q$,
we get the minimal value of energy (38)
$$
\tilde{E}_{min}=\displaystyle\frac{2}{3}v\sqrt{\left(Q+v^2\right)
\left(h^2Q+4m^2v^2\right)}
\eqno(40)
$$
which arises at $\beta$ equal to
$$
\beta_0=\displaystyle\frac{v}{2}\sqrt{\frac{h^2Q+4m^2v^2}{Q+v^2}}.
\eqno
$$
At value of $Q$ equal to one of the exact solution
we get $\beta_0=\displaystyle\frac{hv}{\sqrt{2}}$
and $\tilde{E}_{min}=E=\displaystyle\frac{2\sqrt{2}}{3}hv^3\left(4\frac{m^2}{h^2}-1\right)$,
as it should be.
As it has already been noted, at values of charge sufficiently
close to $Q$ (33), the energy (40) of the new variational function
is less than both the energy value (18) of configuration (15)-(16)
and the energy value (13) of the "kink+plane wave" configuration.
About behavior of the expression (40) at arbitrary value of charge
it is possible to say following.\\
$\bullet$ At  $Q\to+\infty$,
$\tilde{E}_{min}=\displaystyle\frac{2}{3}hvQ+O(Q^{\frac{1}{2}})$,
consquently, at charges greater than $Q^{cr}_1$
(the value $Q^{cr}_1$ can be found from condition of equality 
$\tilde{E}_{min}(Q^{cr}_1)$ (40) and $E_{min}(Q^{cr}_1)$ (18)),
the variational configuration (15)-(16) becomes more energetically
favorable than (36)-(37); so, at large values of charge the 
usual "Q-ball" substitution turns to be more energetically preferable.\\
$\bullet$ At $Q<Q^{cr}_1$, in dependence on ratio $\rho=\displaystyle\frac{h}{m}$
of the Lagrangian constants, two cases are possible.\\
$\star$ Condition of stability of the configuration
with respect to the decay
into "kink + plane wave" is satisfied at any
$Q>0$. It takes place
at $3-\sqrt{5}\le\rho\le 3+\sqrt{5}$. Typical dependence of energy
from charge for "kink+plane wave" and for the two variational configurations
is plotted in Fig.~3a.\\
$\star$ Equality of energy values for the variational and "kink+plane wave"
configurations $\tilde{E}_{min}(Q)=\tilde{E}^0(Q)$
takes place at some
$Q>0$, that corresponds to $\rho>3+\sqrt{5}$ or
$0<\rho<3-\sqrt{5}$. 
In this case, the configuration (36)-(37) at charges
$Q$, smaller than
$Q^{cr}_2=\displaystyle\frac{4}{5}v^2(4/\rho^2-6/\rho+1)$,
is less energetically preferable than "kink+plane wave".
Because of that we can not conclude
about existence of stable localized
solutions at $Q<Q^{cr}_2$. Typical location of the curves on the plot
of energy dependence on charge in this case is presented in Fig.~3b.

     However, in agreement with the conclusion made above,
the energy value (32) of the exact solution is less than value (34)
of the energy of "kink+plane wave" configuration with
corresponding
charge (33) at any $0<\rho\le \sqrt{2}$. So, the variational
function (36)-(37) is an exact solution at charge values $Q=0$
and $Q=2v^2(2/\rho^2-1)$. At the same time, Eqs. (36)-(37)
at any $Q>Q^{cr}_2$ (or at any $Q\ge 0$, in dependence on the value of
$\rho$) give to the energy functional
the smaller value than
the "kink+plane wave" configuration does, and at any value of charge 
such that $Q<Q^{cr}_1$ (36)-(37) leads to the value of the energy functional
smaller than the variational configuration (15)-(16) does.

     Thus, we have improved the variational function for charges
smaller than the critical one $\tilde{Q}^{cr}$ for the functions
(15)-(16). Besides that the new variational function coincides with
the exact solution
(29)-(30) at the corresponding value of charge. The exact solution found
(29)-(30) and improved variational procedure (36)-(37) lead to the
conclusion that the topological Q-balls
can exist in much wider range of 
charge values than it follows from the standard variational
procedure which is usually applied to Q-balls.

\section{Conclusion}

   So, we have demonstrated that a wall-type one-dimensional 
solution (kink) for the scalar Higgs field is able to bind
the complex scalar field with U(1) charge; the value of
charge $Q$ bound by the wall can vary in a wide range.

     The ability of the topological solution to bind large U(1) 
charge is the main new result of our work. This effect
should appear in the problem of scattering of a particle on
a domain wall. This problem was considered first by
Voloshin \cite{Vol1}. Two different problems were considered in
that paper.
The first one was the investigation of scattering of a
plane wave of the Higgs field on a kink of the same field. The
scattering turns to be unreflective in this case.
In the second problem, an interaction between the Higgs field
and fermionic field was considered. The problem of scattering
of a fermion on a kink was also reduced to the
scattering on given potential created by the kink field.
The formulation of this problem following the work \cite{Vol1}
in our case is the investigation of scattering problem for
Eq. (4) at given potential $\phi(x)$.
With substitution of $\phi(x)$ in the form of (8), the problem
is reduced to the well-known problem of scattering of a particle
in the external potential $V(x)\sim -1/\cosh^2{(mvx)}$.
It is interesting that at $h\to 0$ the reflection coefficient
for this potential goes to one at momentum $k\to 0$ \cite{Landau}.
Such behavior of the reflection coefficient at $h\to 0$ means
that the limit of weak coupling does not work.
It is not accidental in this case because in one dimension
a bound state in an attractive potential always exists.

     It should be noted that the problem of scattering
of a particle on the kink can not be reduced to the problem of
scattering on a given potential in general case.
Indeed, we have to consider Eq. (4) which
defines the behavior of $\xi$ at given $\phi(x)$
together with Eq. (5) for the aim of 
investigating how does $\phi(x)$ vary in
the field of the plane wave $\xi(x)$. How it may be seen
from (5), at substituting $\xi(x)$ as the plane
wave the vacua of $\phi$ and the mass of excitation
of $\phi$ over the vacua are changing. Thus, the problem
of scattering of particles in the field of kink is
self-consistent only if these particles are the small
deviations of the field $\xi$ over its vacuum. Moreover, even a
bound state of $\xi$ in the potential of the
kink can be obtained by neglecting back $\xi$ to $\phi$ 
influence formally only if this state is bound
strongly enough. It can be understood qualitatively
by taking into account that the wave function of a
weakly bound state falls down slowly at $x\to\pm\infty$.
Because of that in (5) the neglected term $2h^2\xi\xi^*\phi$
at $\phi(x)=v\tanh{(mvx)}$ becomes exponentially large
at $x\to\pm\infty$ in comparison to the term
$(\phi^2-v^2)\phi\sim\cosh^{-2}{(mvx)}$.
In case of existence of the only bound state in potential
$-h^2v^2/\cosh^2{(mvx)}$
the wave function of this state is
\cite{Landau}:
$$
\xi=\displaystyle\frac{e^{i\omega t}}{\cosh^s{(mvx)}}, \quad \mbox{где}
\quad s=\displaystyle\frac{1}{2}\left(-1+\sqrt{1+4\rho^2}\right).
$$
For the term $2h^2\xi\xi^*\phi$ not to be exponentially large 
at large $x$ in comparison to the other terms in (5),
it is necessary to be $s\ge 1$ from where we get $\rho\ge\sqrt{2}$.
So, not the only scattering problem, but also the problem of a
weakly bound $\xi$ particle in the kink field should be solved
with taking into account the reverse influence of the field $\xi$
to the field $\phi$.

     The possibility of existence of domain walls carrying U(1) charge
can be immediately related to the problem of domain bubble collapse.
The problem of shrinking of domain bubble was investigated
in detail and in different aspects starting from the pioneer
work of Ya.~B.~Zeldovich, I.~Yu.~Kobzarev, and L.~B.~Okun \cite{Zeld}
(see also, e.g.,\cite{Kuzmin}).
Usually, the evolution of the spherically symmetric bubbles of the
$\lambda\phi^4$ theory having the radial field profile corresponding
to that of kink, being considered. As it is known, such bubbles are
to shrink and there is no stationary domain structure in
the $\lambda\phi^4$ theory.
The possibility of stabilization of the domain bubble by filling
with the charge was considered starting from Refs.\cite{Viciarelly,Drell}.
In particular, the possibility of bubble stabilization by filling it
with quarks has been discussed by Bardeen, Chanowitz, Drell et al.
\cite{Drell}.
The charged wall solution found in this our work could also be used
for searching of a stable U(1) charged domain area. Here should be
emphasized that a vacuum domain surrounded by charged domain wall
could evolve into a Q-ball. We are going to return to consideration
of evolution of such configurations.

     The U(1) charged solutions like ones found in this our work
could exist in some more general models. In particular,
a "hedgehog" solution for the Higgs field triplet interacting
with the complex field with U(1) symmetry is of interest.

    The question of stability of the found exact solution and the
variational solutions has been considered on the level of stability
with respect to decay into uncharged kink and plane waves.
The question of stability of the exact solution with respect to small
deformations will be discussed in detail in a separate publication.

     The authors would like to thank N.~A.~Voronov, N.~B.~Konyukhova
and S.~V.~Kurochkin for useful discussions.

     This work was supported in part by the Russian Foundation for
Basic Research, grant number 00-15-96562.

\newpage

\section*{Figure captions}

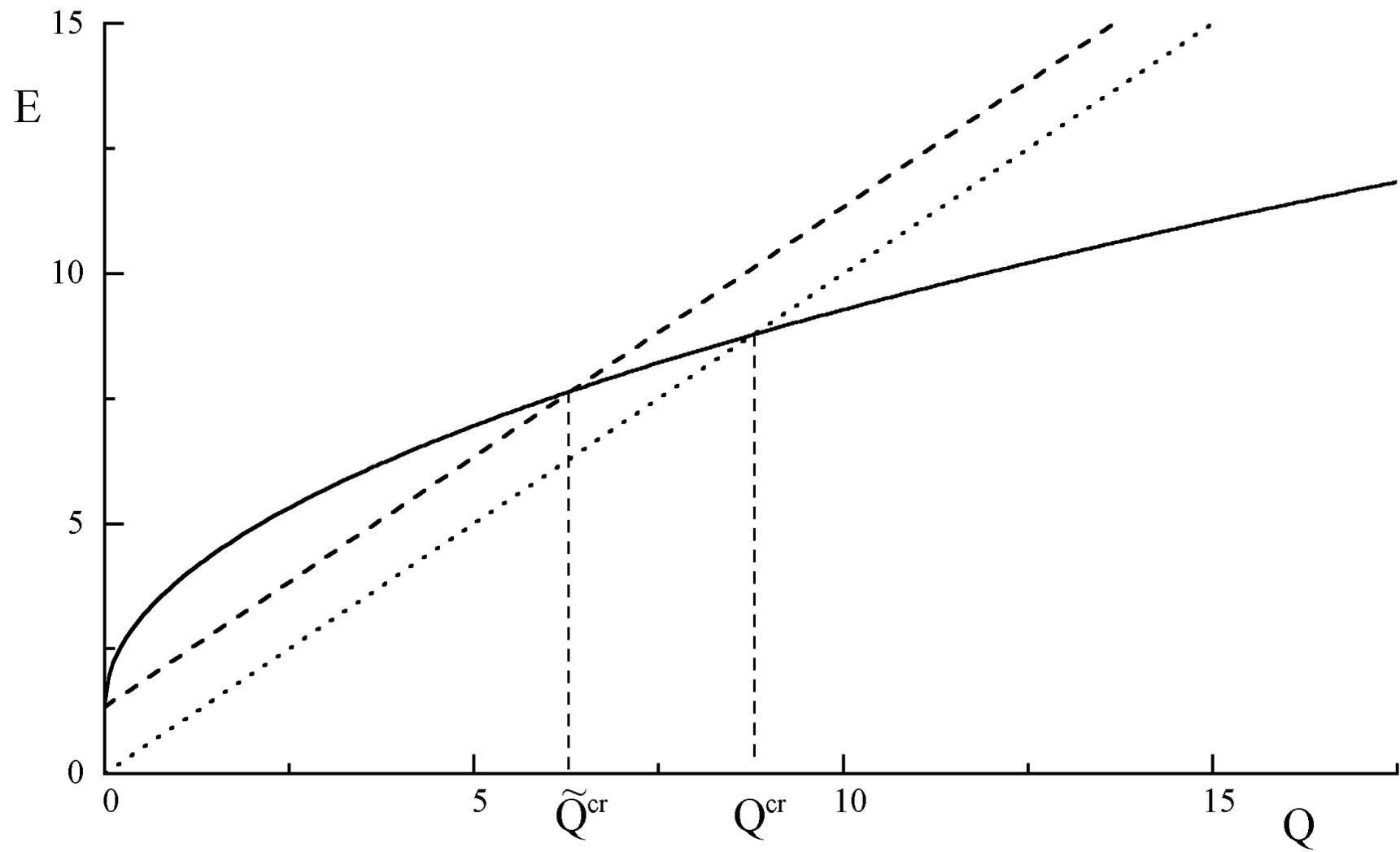
{\bf Fig. 1.} The dependences of the energy on the charge
for the "plane wave" configuration (dotted curve), for the
"kink+plane wave" configuration (dashed curve) and for the
variational configuration (15)-(16) (solid curve).
$Q^{cr}$ is the critical value of the charge for the
non-topological configuration;
$\tilde{Q}^{cr}$ is the critical value of the charge for
the topological configuration. Curves are plotted at $h=m=1$;
$v=1$.

\bigskip

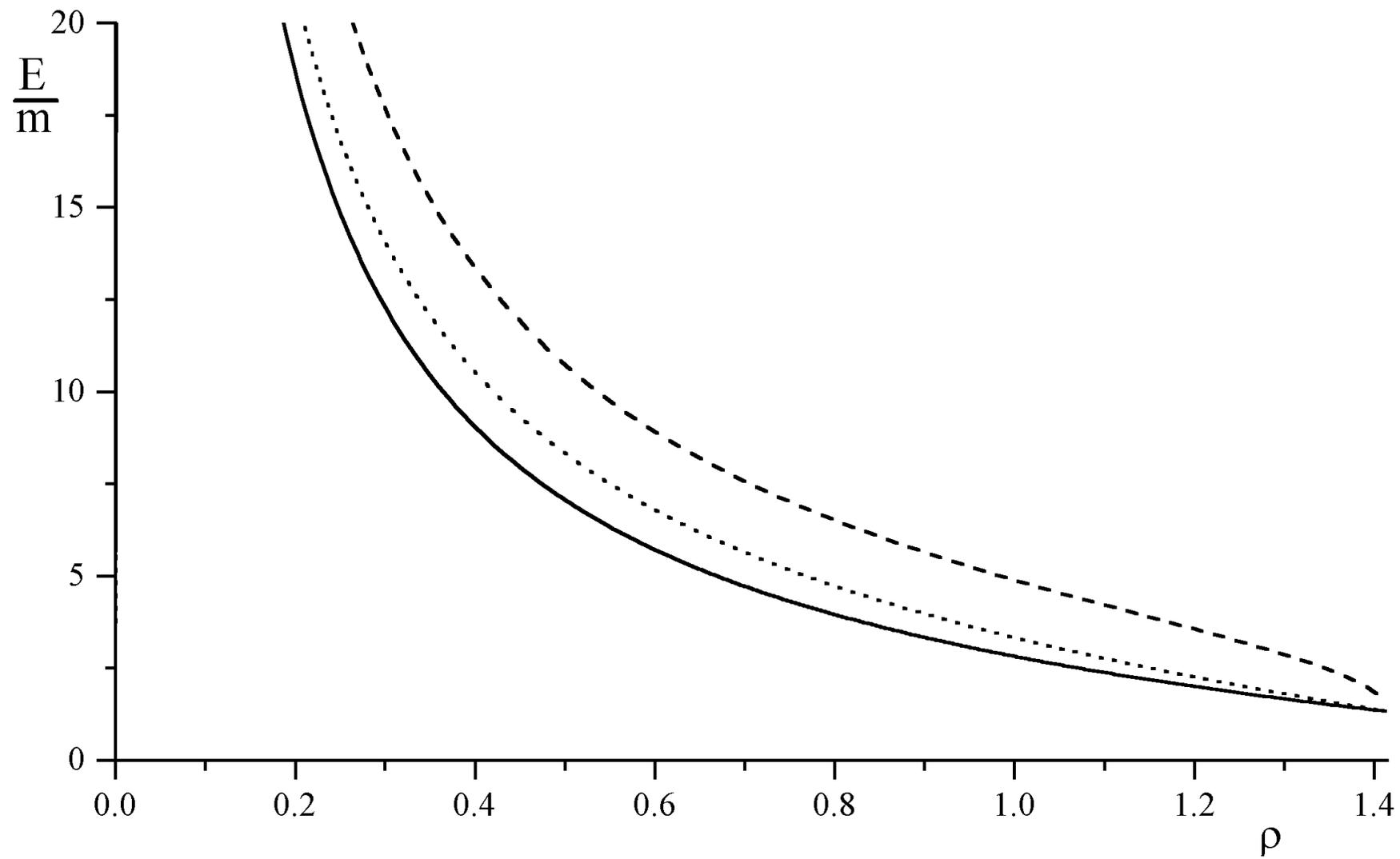
{\bf Fig. 2.} The dependences of the ratio $E/m$ versus
$\rho$ for the exact solution (solid curve),
for the "kink+plane wave" configuration (dotted curve)
and for the variational configuration (15)-(16) (dashed curve).
The charge of the last two configurations equals to $Q$ (33).
Curves are plotted at $v=1$.

\bigskip

{\bf Fig. 3.} The dependences of the energy on the charge
for the variational configurations (15)-(16) (dashed curve),
(36)-(37) (solid curve), and for the "kink+plane wave" configuration
(dotted curve), corresponding to:\\
a) $\rho=1$ ($h=m=1$), $v=1$;\\
b) $\rho=\sqrt{0.1}=0.31622\dots$ ($h=\sqrt{0.1}, m=1$), $v=1$.

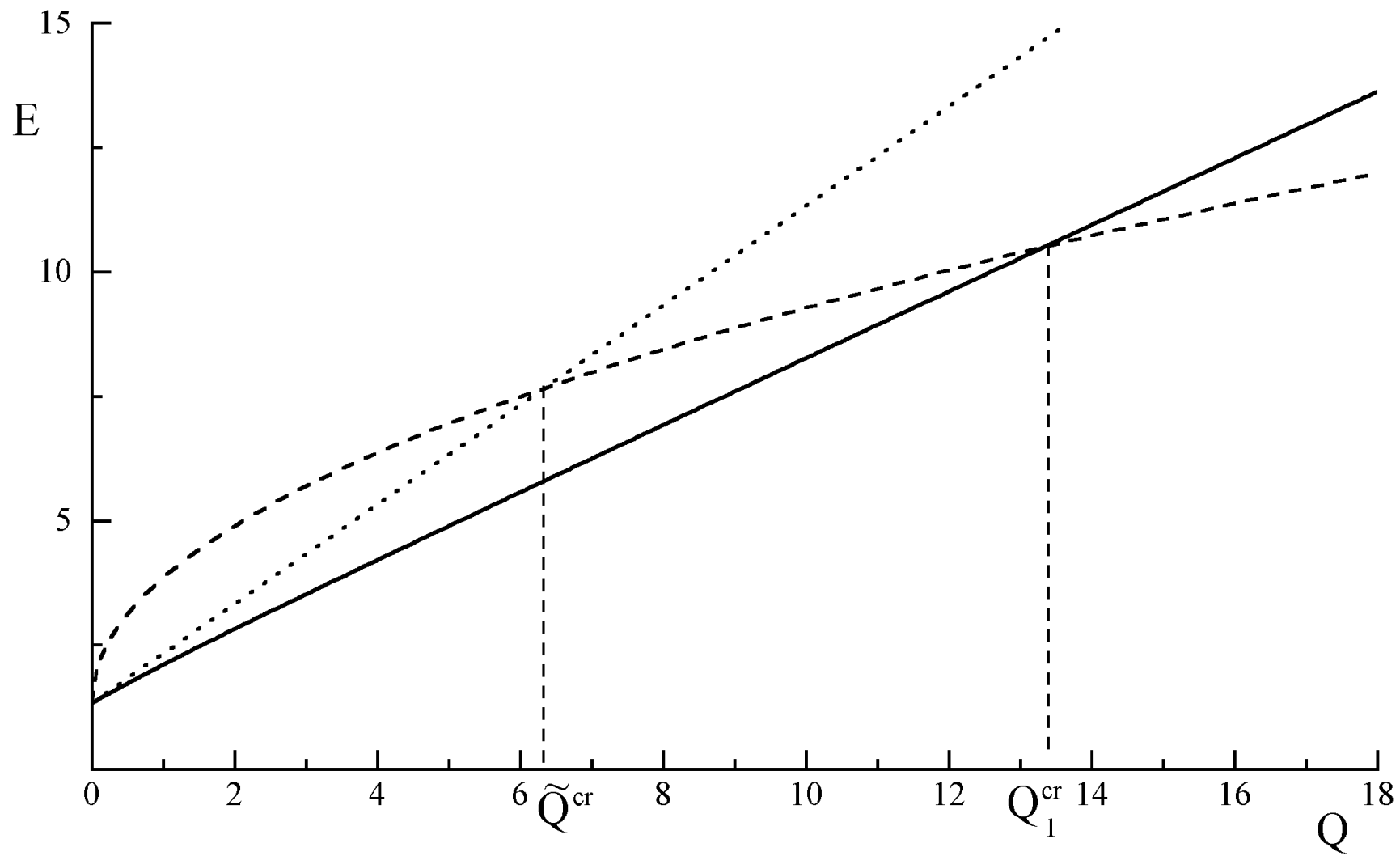
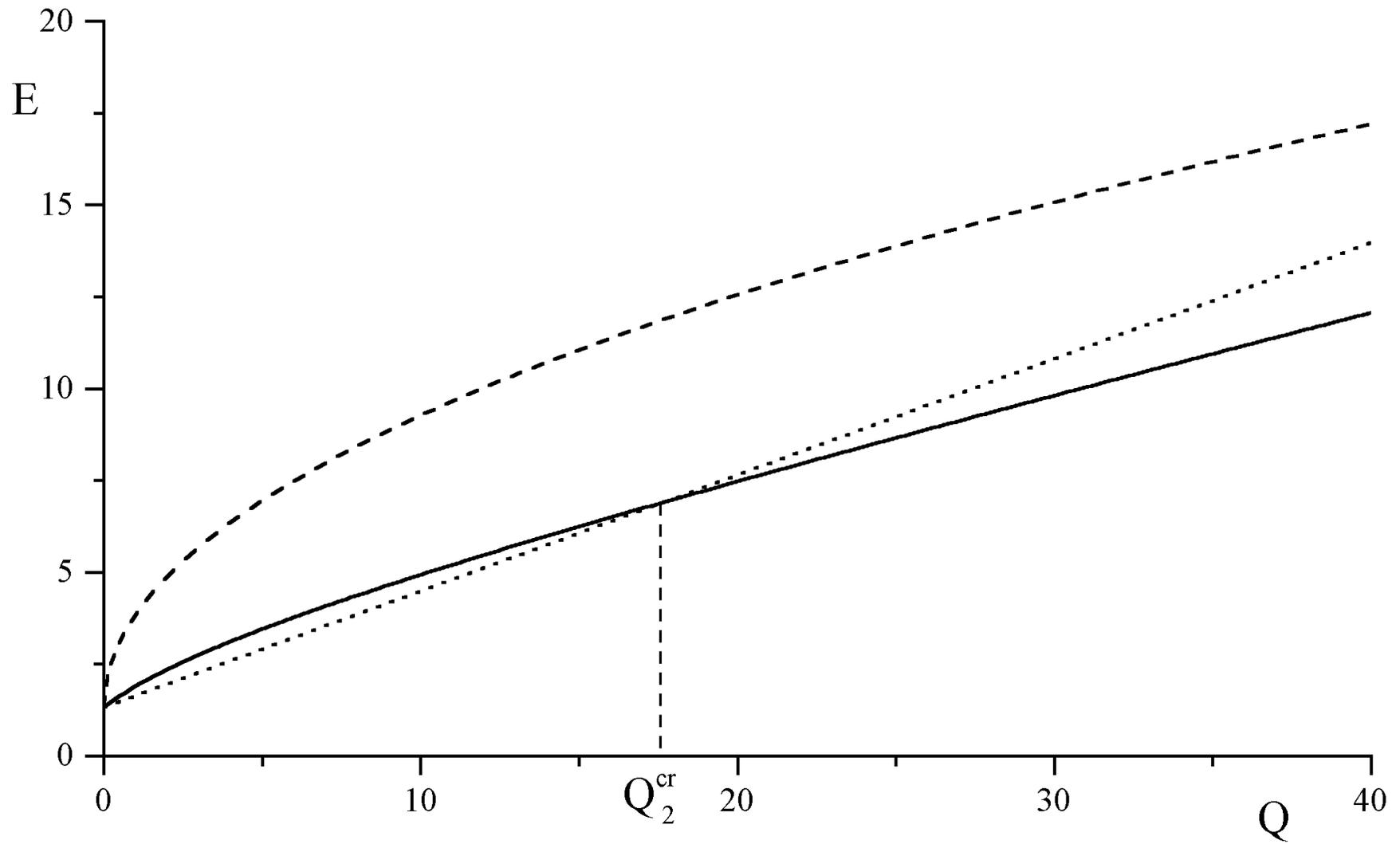
\end{document}